\documentclass[10pt]{article}
\usepackage{graphicx}

     \usepackage{bm}

\def\Title#1{\begin{center} {\Large #1 } \end{center}}
\def\Author#1{\begin{center}{ \sc #1} \end{center}}
\def\Address#1{\begin{center}{ \it #1} \end{center}}

\newcommand\pubblock{\rightline{\begin{tabular}{l} Proceedings of the Second Annual LHCP\\
         \pubdate  \end{tabular}}}

\newenvironment{Abstract}{\begin{quotation} \begin{center} 
             \large ABSTRACT \end{center}\bigskip 
      \begin{center}\begin{large}}{\end{large}\end{center} \end{quotation}}

\newenvironment{Presented}{\begin{quotation} \begin{center} 
             PRESENTED AT\end{center}\bigskip 
      \begin{center}\begin{large}}{\end{large}\end{center} \end{quotation}}

\def\beq{\begin{equation}}
\def\eeq#1{\label{#1}\end{equation}}
\def\eeqn{\end{equation}}

\def\beqa{\begin{eqnarray}}
\def\eeqa#1{\label{#1}\end{eqnarray}}
\def\eeqan{\end{eqnarray}}

\let\bar=\overbar

\def\L{{\cal L}}

\def\Dslash{\not{\hbox{\kern-4pt $D$}}}
\def\dslash{\not{\hbox{\kern-2pt $\del$}}}

\def\msb{{\bar{\ssstyle M \kern -1pt S}}}

\usepackage{color}

\newcommand\pubdate{\today}

\def\affiliation{
           Institute of Nuclear Physics PAN, PL-31-342 Cracow, Poland \\
           and University of Rzesz\'ow, PL-35-959 Rzesz\'ow, Poland  }

\def\support{\footnote{This work was supported in part by the Polish grants
DEC-2011/01/B/ST2/04535 and  DEC-2013/09/D/ST2/03724 as well as by the Centre for
Innovation and Transfer of Natural Sciences and Engineering Knowledge in
Rzesz{\'o}w.}}

\begin{document}

\large
\begin{titlepage}
\pubblock

\vfill
\Title{  PRODUCTION OF CHARMED MESON-MESON PAIRS AT THE LHC: SINGLE- VERSUS DOUBLE-PARTON SCATTERING MECHANISMS}
\vfill

\Author{ RAFAL MACIULA and ANTONI SZCZUREK \support }
\Address{\affiliation}
\vfill
\begin{Abstract}
  We discuss hadroproduction of open charmed mesons ($D^0$, $D^{\pm}$, $D^{\pm}_{S}$) at the LHC energy $\sqrt{s} = 7$ TeV. The cross section for inclusive production of $c \bar c$ pairs is calculated within the $k_{\perp}$-factorization (or high-energy factorization) approach which effectively includes higher-order corrections. Results of the $k_{\perp}$-factorization approach are compared to NLO parton model predictions. The hadronization of charm quarks is included with the help of the Peterson fragmentation functions. Inclusive differential distributions in (pseudo)rapidity and transverse momentum for several charmed mesons  are calculated and compared to recent results of the ALICE, ATLAS and LHCb experiments. We also take into consideration a mechanism of double charm (two pairs of $c \bar c$) production within a simple
formalism of double-parton scattering (DPS). Surprisingly for LHC energies the DPS
cross sections are found to be larger than those from the standard SPS mechanism. We compare our predictions for $DD$ meson-meson pair production with recent measurements of the LHCb collaboration, including correlation observables. Our calculations clearly confirm the dominance of DPS in the
production of double charm, however some strength seems to be still lacking.
Possible missing contribution from the so-called single-ladder-splitting DPS mechanism is also discussed.

\end{Abstract}
\vfill

\begin{Presented}
The Second Annual Conference\\
 on Large Hadron Collider Physics \\
Columbia University, New York, U.S.A \\ 
June 2-7, 2014
\end{Presented}
\vfill
\end{titlepage}
\def\thefootnote{\fnsymbol{footnote}}
\setcounter{footnote}{0}
%

\normalsize 


\section{Introduction}
\label{intro}

Inclusive distributions for different species of pseudoscalar $D$ and vector $D^{*}$ open charm meson have been measured recently by ATLAS \cite{ATLASincD}, ALICE \cite{ALICEincD} and LHCb \cite{LHCbincD} experiments. The LHCb group has also performed more exclusive studies of correlation observables for $D\overline{D}$ pairs in the forward rapidity region $2 < y < 4$ \cite{LHCb-DPS-2012}.

Usually, in numerical studies of heavy quark production the main efforts concentrate on inclusive distributions.
Improved schemes of standrad pQCD NLO collinear approach are state of art in this
respect. On the other hand, the $k_t$-factorization approach is commonly used as a very efficient tool for more exclusive studies of kinematical correlations between produced particles (see e.g.~\cite{Maciula:2013wg} and references therein). In this approach the transverse momenta of incident partons are taken into account and their emission is described by the so-called unintegrated gluon distribution functions (UGDFs). This allows to construct different correlation distributions which are strictly related to the transverse momenta of incident particles. 

It was concluded recently that the cross section
for $c \bar c c \bar c$ production at the LHC may be very large
due to mechanism of double-parton scattering (DPS) \cite{Luszczak:2011zp}. 
Meanwhile, the LHCb collaboration measured unexpectedly large cross section for the production of $DD$ meson-meson pairs at 
$\sqrt{s}$ = 7 TeV \cite{LHCb-DPS-2012}.
Those data sets for double open charm production have been studied differentially only
within the $k_{\perp}$-factorization approach, where several correlation observables useful to identify the DPS effects have been carefully discussed. In order to draw definite conclusions about the DPS effects in double charm production it is necessary to estimate precisely contribution to $c \bar c c \bar c$ final state from the standard mechanism of single-parton scattering \cite{HMS2014}.

\section{Inclusive single charmed meson production}
\label{sec-1}

The cross section for the production of a pair of charm quark -- 
charm antiquark within the $k_t$-factorization approach can be written as:
\begin{eqnarray}
\frac{d \sigma(p p \to c \bar c X)}{d y_1 d y_2 d^2 p_{1t} d^2 p_{2t}} 
\! = \! \frac{1}{16 \pi^2 {\hat s}^2} \! \int \! \frac{d^2 k_{1t}}{\pi} \frac{d^2 k_{2t}}{\pi} \overline{|{\cal M}^{off}_{shell}|^2}\! \times \! \delta^2 \! \left( \vec{k}_{1t}\! +\! \vec{k}_{2t}\! -\! \vec{p}_{1t}\! -\! \vec{p}_{2t}
\right)
\! {\cal F}_g(x_1,k_{1t}^2,\mu^2) {\cal F}_g(x_2,k_{2t}^2,\mu^2).
\end{eqnarray}

The essential components in the formula above are off-shell matrix elements for $g^{*}g^{*} \rightarrow c \;\bar{c}$ subprocess
and unintegrated (transverse momentum dependent) gluon distributions (UGDF). The relevent matrix
elements are known and can be found, e.g in Ref.~\cite{CCH91}. 
The unintegrated gluon distributions are functions of
longitudinal momentum fraction $x_1$ or $x_2$ of gluon with respect to its parent nucleon and of gluon transverse momenta $k_{t}$.
Some of them depend in addition on the factorization scale $\mu$.
In contrast to the collinear gluon distributions (PDFs) they differ
considerably among themselves. 
One may expect that they will lead to different
production rates of $c \bar c$ pairs at the LHC. Since the production of charm quarks
is known to be dominated by the gluon-gluon fusion, the charm production in hadronic reactions at the LHC
can be used to verify the quite different models of UGDFs. 

The hadronization of heavy quarks is done
with the help of fragmentation functions technique, using
the standard Peterson model \cite{Peterson}, normalized to
the proper branching fractions from Ref.~\cite{Lohrmann2011}.
%
\begin{figure}[!h]
\centering
\includegraphics[width=6.cm]{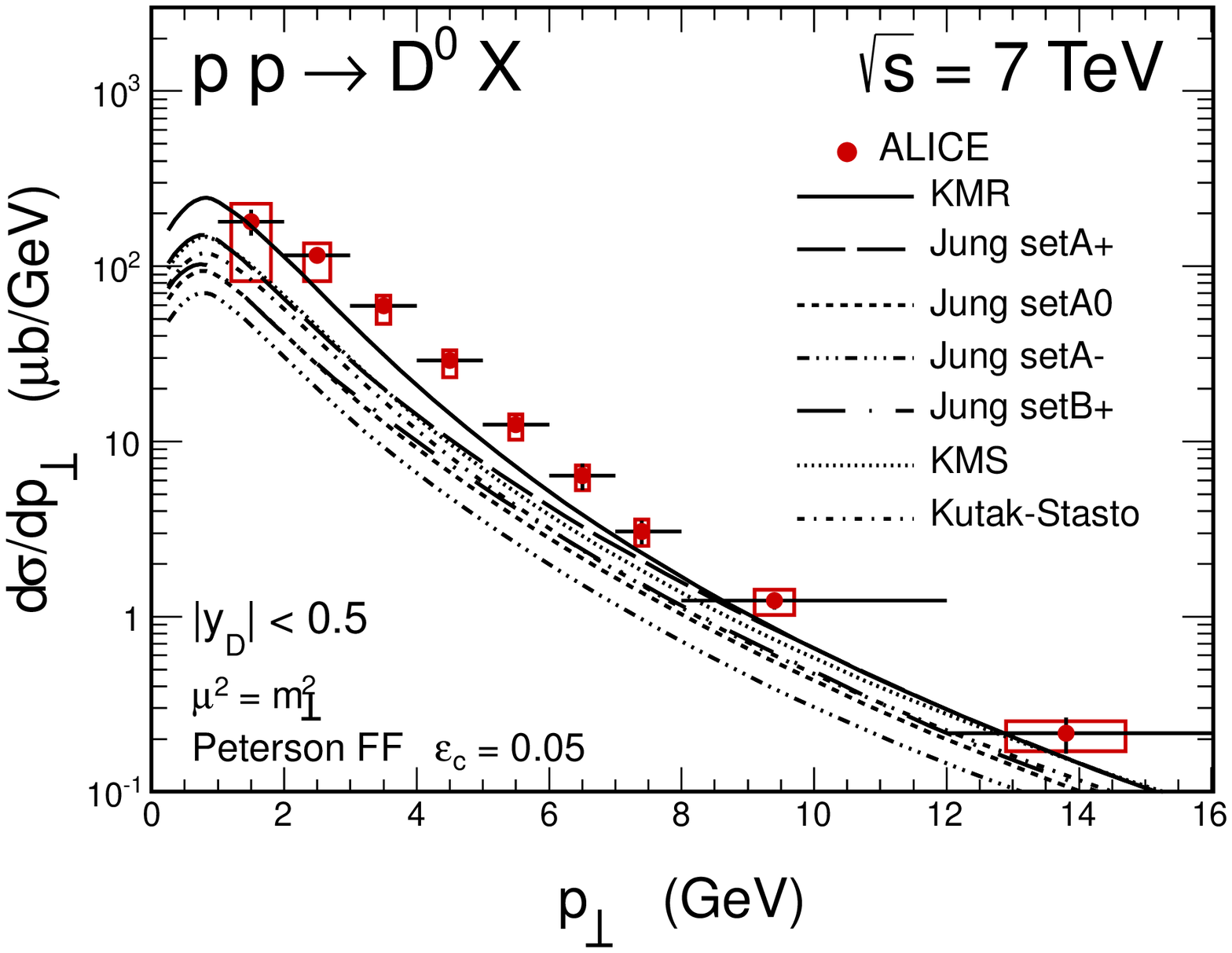}
\includegraphics[width=6.cm]{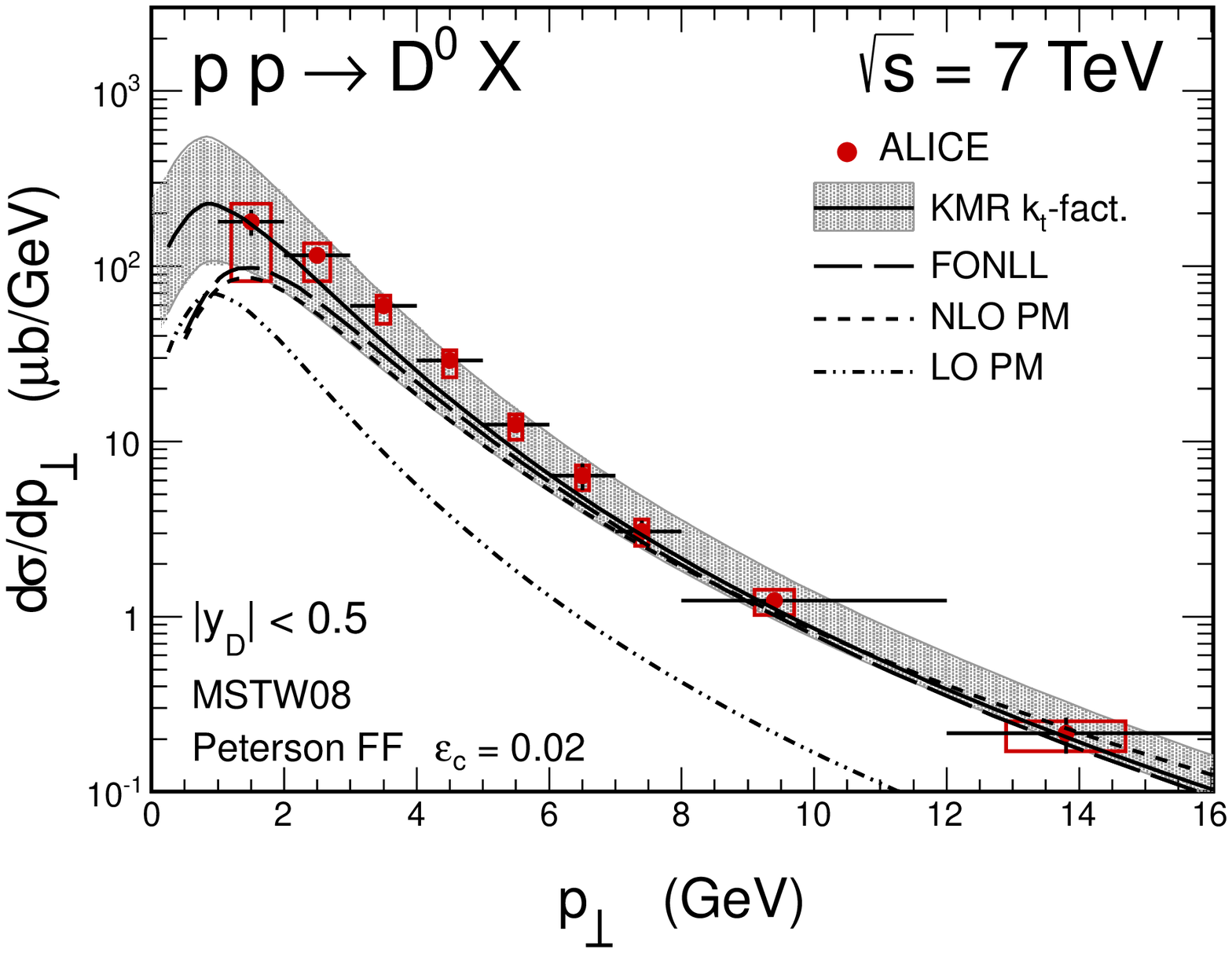}
   \caption{
\small Transverse momentum distribution of $D^0$ meson for different UGDFs (left) and for the KMR UGDF compared to the pQCD collinear calculations (right) together with the ALICE data.
}
 \label{fig:pt-single-alice}
\end{figure}

\begin{figure}[!h]
\centering
\includegraphics[width=6.cm]{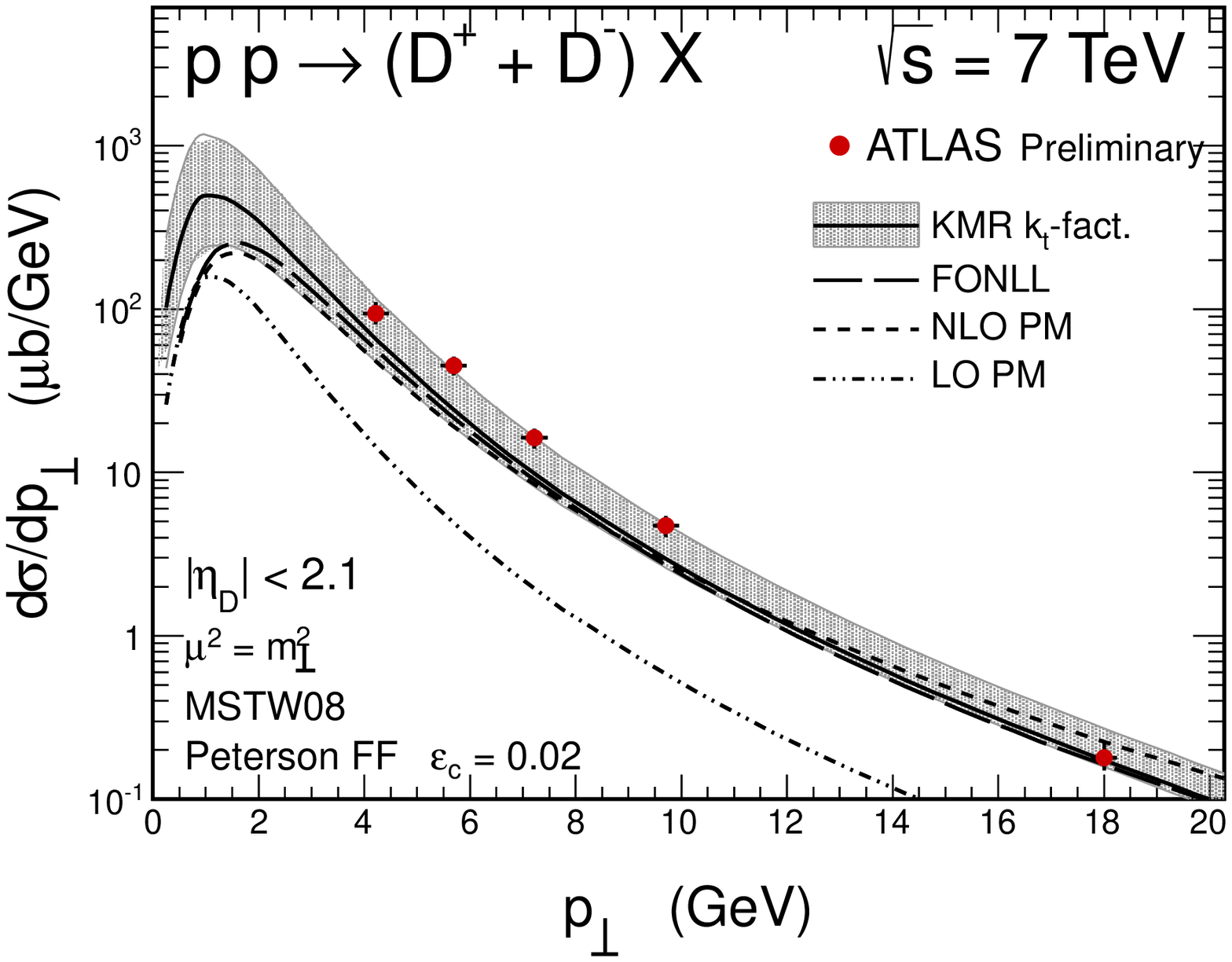}
\includegraphics[width=6.cm]{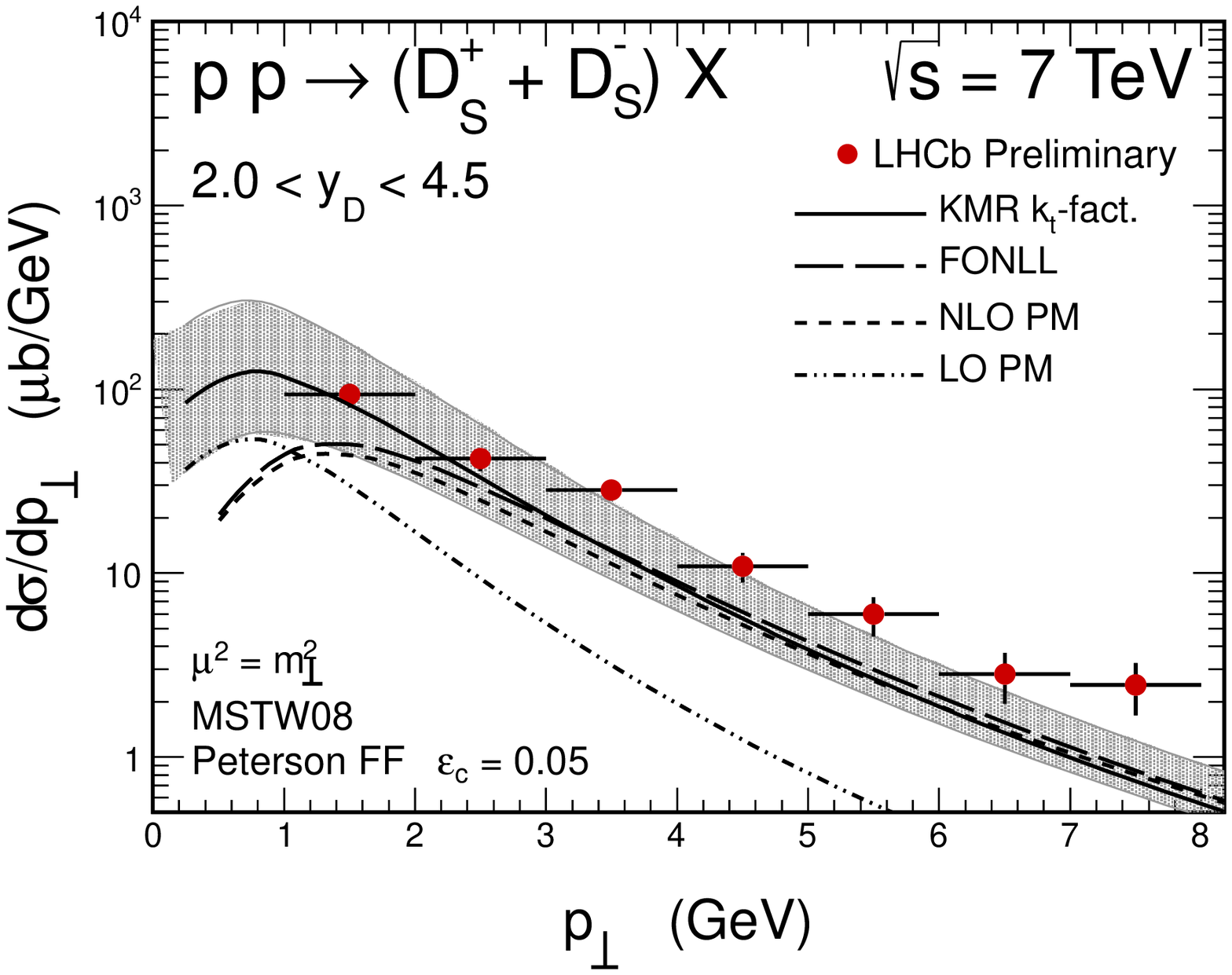}
\includegraphics[width=6.cm]{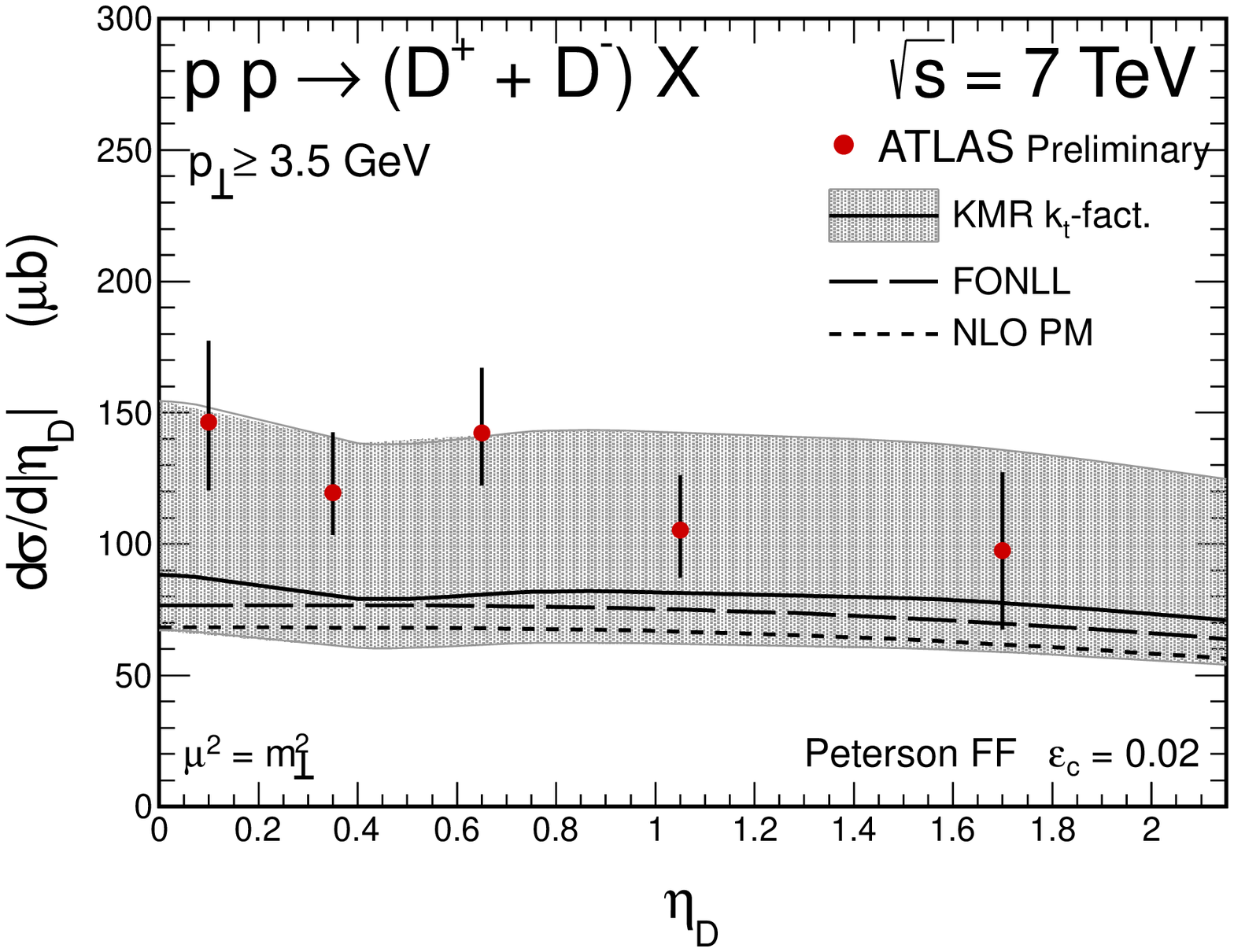}
\includegraphics[width=6.cm]{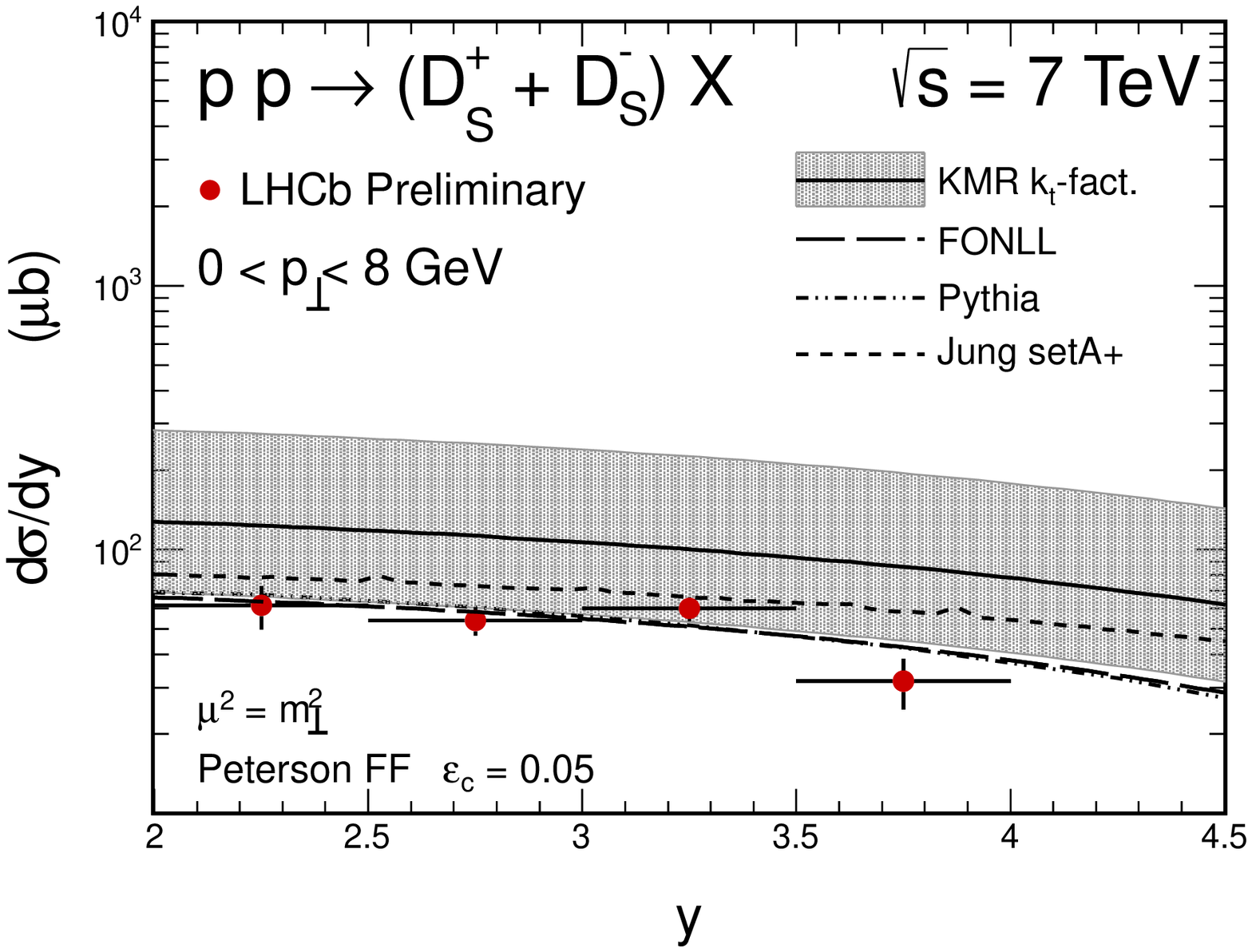}
   \caption{
\small Transverse momentum (top) and (pseudo)rapidity (bottom) distribution of $D^{\pm}$ meson for the ATLAS experiment (left) and of $D^{\pm}_{S}$ meson for the LHCb experiment (right). Together with our predictions for the KMR UGDF (solid line with shaded band). Results of pQCD collinear approaches are also shown.
}
 \label{fig:pt-single-atlas-lhcb}
\end{figure}

In Fig.~\ref{fig:pt-single-alice} we show transverse momentum distribution
of $D^0$ mesons. In the left panel we present results for different UGDFs
known from the literature. Most of the UGDFs applied here fail
to describe the ALICE data. The KMR UGDF \cite{KMR} provides the best description
of the measured distributions. Therefore in the following we shall concentrate only on
the calculations based on the KMR UGDF. In the right panel we show the uncertainties of our predictions due to the charm quark mass $m_c = 1.5 \pm 0.3$ GeV and related with the renormalization and factorization scales $\mu^{2} = \zeta m_{t}^{2}$, where $\zeta \in (0.5; 2)$. The gray shaded bands
represent these both sources of uncertainties summed in quadrature. In addition, the results of relevant calculations
within the LO and NLO collinear approach are presented for comparison. The $k_{t}$-factorization approach with the KMR UGDF is consistent
with the NLO collinear calculations.  

The left panels of Fig.~\ref{fig:pt-single-atlas-lhcb} shows transverse momentum (top) and pseudorapidity (bottom) distributions of charged pseudoscalar $D^{\pm}$ mesons measured by the ATLAS experiment. Overall situation is very similar as for the ALICE case except of the agreement with the experimental data points, which is somewhat worse here. Only the very upper limit of the KMR result is consistent with the ATLAS data. This is also true for the other standard collinear NLO pQCD approaches. The worse description of the ATLAS data may be caused by much broader range of pseudorapidities than in the case of the ALICE detector. Potentially, this can be related to double-parton scattering (DPS) effects \cite{Maciula:2013kd}.
 
Recently the LHCb collaboration presented first results for
the production of different $D$ mesons in in the forward rapidity region $2 < y < 4.5$.
In this case one can test asymmetric configuration of gluon longitudinal
momentum fractions: $x_1 \sim$ 10$^{-5}$ and $x_2 >$ 10$^{-2}$.
Standard collinear gluon distributions as well as unintegrated ones were never tested at such small values of $x$.
In the right panels of Fig.~\ref{fig:pt-single-atlas-lhcb} we present distributions corresponding to the ATLAS case but for $D_s^{\pm}$ meson.
The main conclusions are the same as for ALICE and ATLAS conditions. Our results with the KMR UGDF within uncertainties are consistent (with respect to the upper limits) with the experimental data and with the NLO collinear predictions.
\section{Single- and double-parton scattering effects in $\bm{DD} $ meson-meson pair production at the LHC}
\label{sec-2}

\begin{figure}[!h]
\centering
\includegraphics[width=4cm]{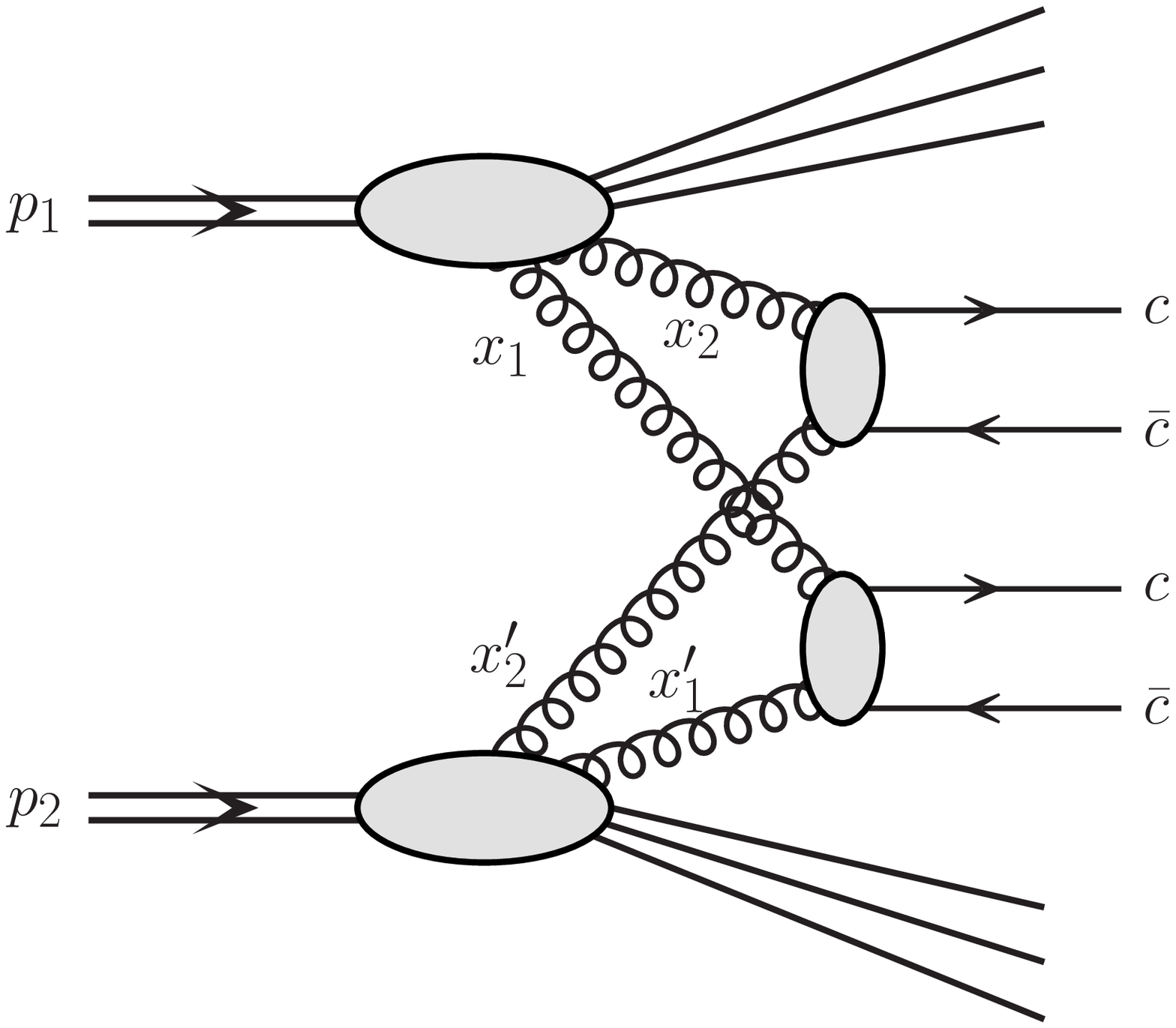}
\includegraphics[width=4cm]{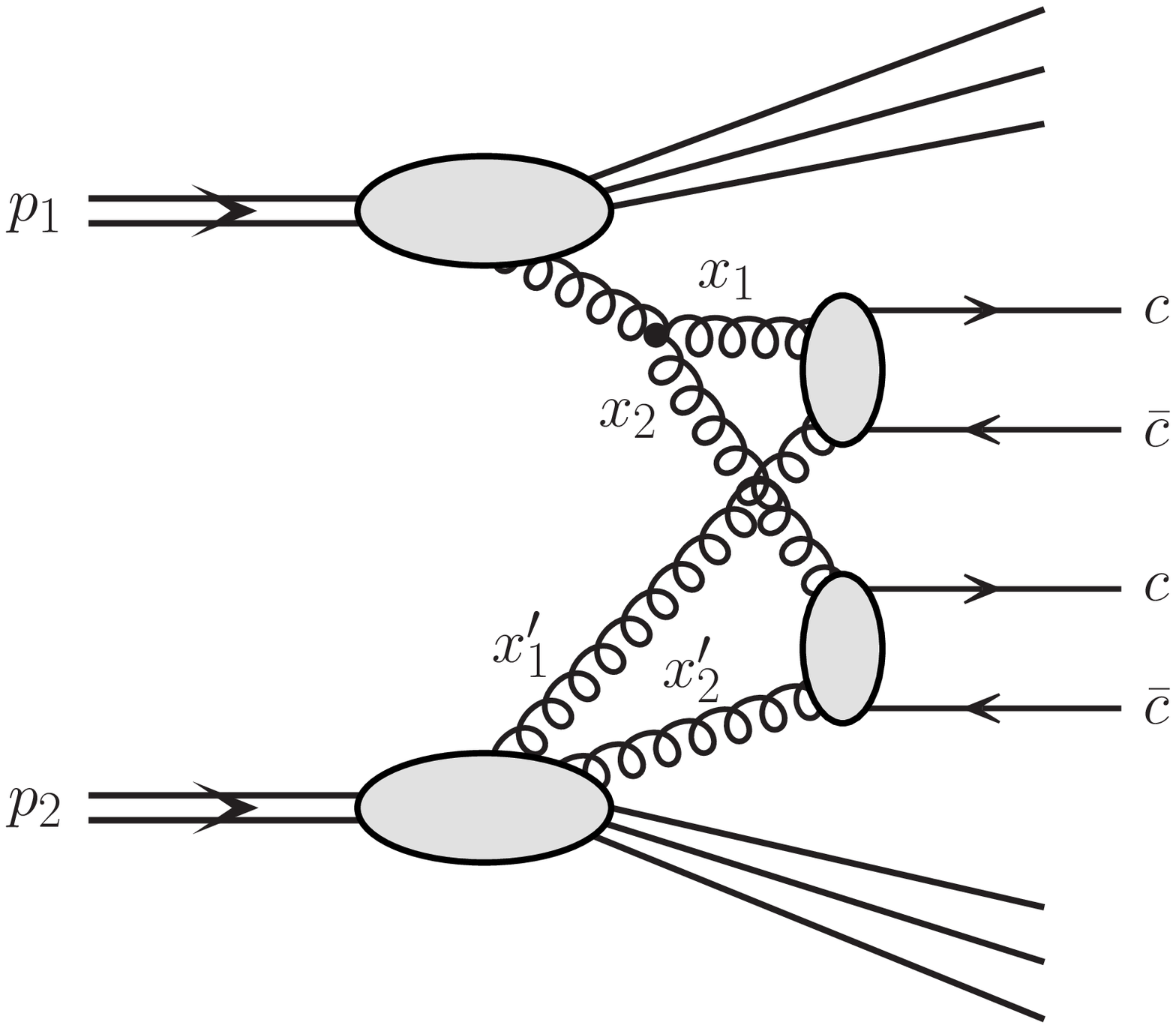}
\includegraphics[width=4cm]{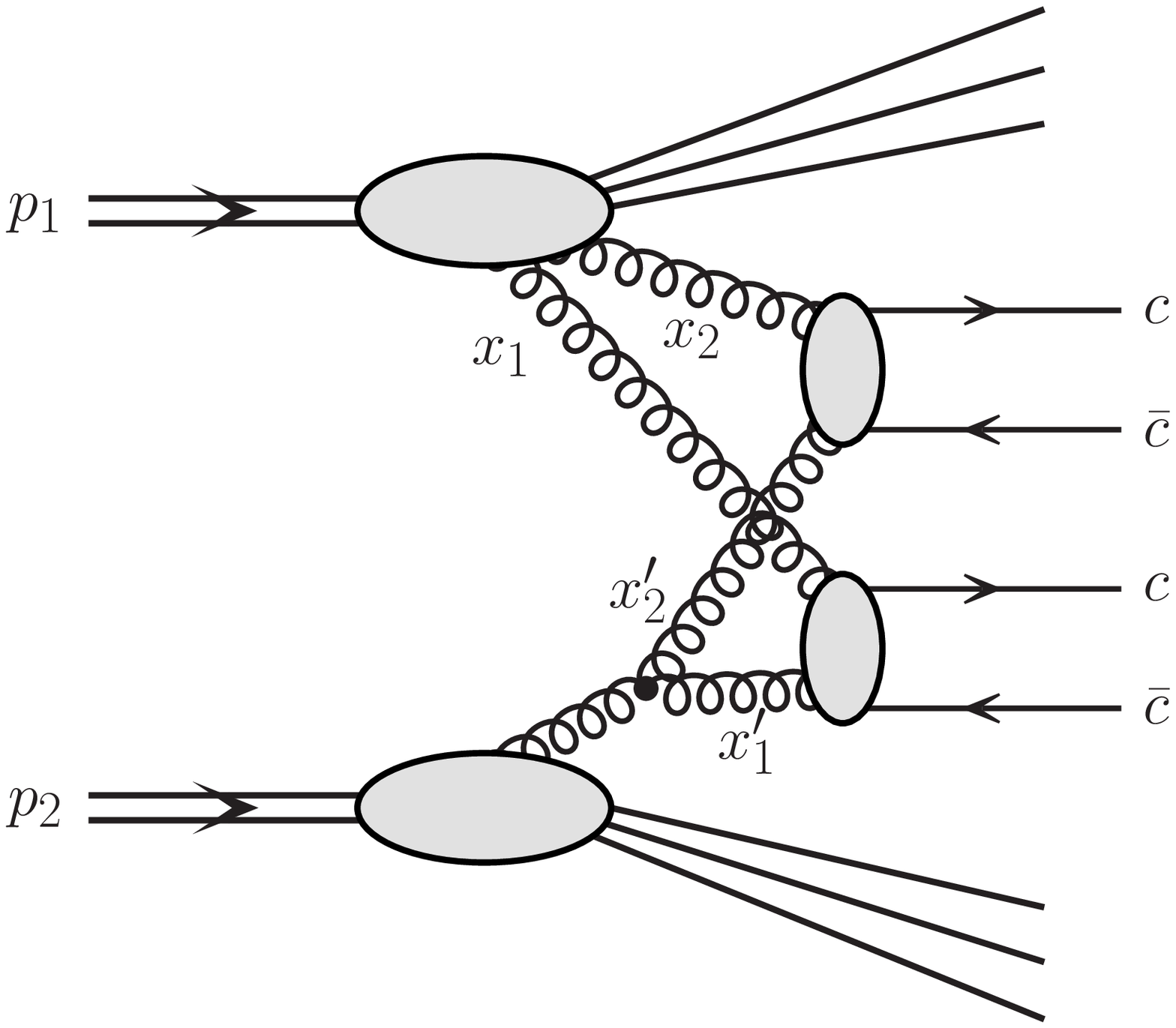}
   \caption{
\small The standard (left) and single-ladder-splitting (middle and right) diagrams for DPS production of $c \bar c c \bar c$.
}
 \label{fig:diagrams_ccbarccbar}
\end{figure}

Production of $c \bar c c \bar c$ (4-parton) final state is particularly
interesting especially in the context of experiments being carried out at the LHC
and has been recently carefully discussed \cite{Luszczak:2011zp,Maciula:2013kd}. 
The double-parton scattering formalism in the simplest form assumes two
independent standard single-parton scatterings (see left diagram of Fig.~\ref{fig:diagrams_ccbarccbar}). Then in a simple probabilistic picture, in the so-called factorized Ansatz, the differential cross section for DPS production of $c \bar c c \bar c$ system in the $k_{\perp}$-factorization approach can be written as:
\begin{eqnarray}
\frac{d \sigma^{DPS}(p p \to c \bar c c \bar c X)}{d y_1 d y_2 d^2 p_{1,t} d^2 p_{2,t} 
d y_3 d y_4 d^2 p_{3,t} d^2 p_{4,t}} = 
\frac{1}{2 \sigma_{eff}} \cdot
\frac{d \sigma^{SPS}(p p \to c \bar c X_1)}{d y_1 d y_2 d^2 p_{1,t} d^2 p_{2,t}}
\cdot
\frac{d \sigma^{SPS}(p p \to c \bar c X_2)}{d y_3 d y_4 d^2 p_{3,t} d^2 p_{4,t}}.
\end{eqnarray}

This formula assumes that the two subprocesses are not correlated and do not interfere. The parameter $\sigma_{eff}$ in the denominator of the above formula
from a phenomenological point of view is a non-perturbative quantity
related to the transverse size of the hadrons and has the dimension of a cross section.
The dependence of $\sigma_{eff}$ on the total energy at fixed scales is rather small and
one expect, that the value should be equal to the total non-diffractive cross
section, if the hard-scatterings are really uncorrelated.
More details of the theoretical framework for DPS mechanism applied here can be found in Ref.~\cite{Maciula:2013kd}.

\begin{figure}[!h]
\centering
\includegraphics[width=6.cm]{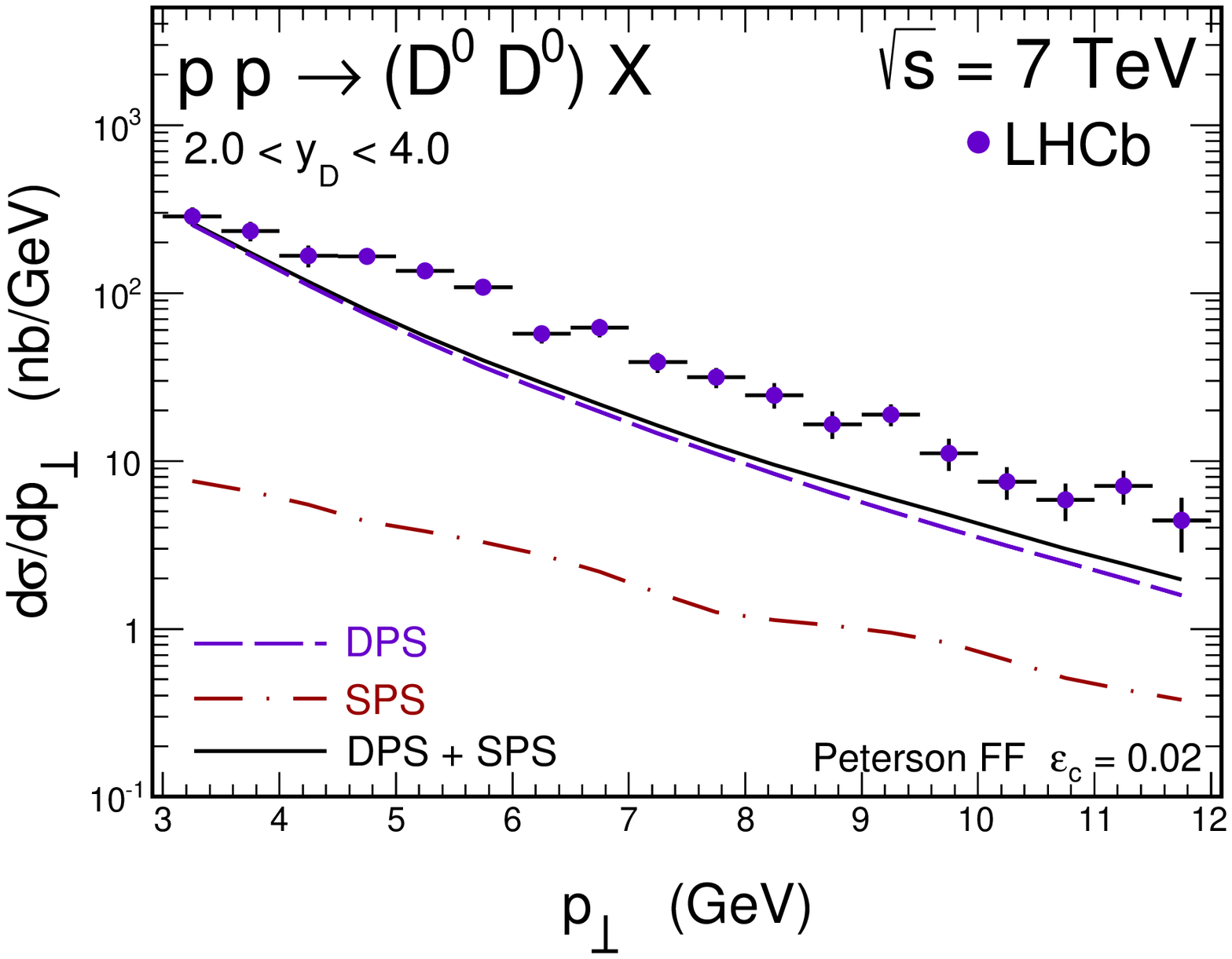}
\includegraphics[width=6.cm]{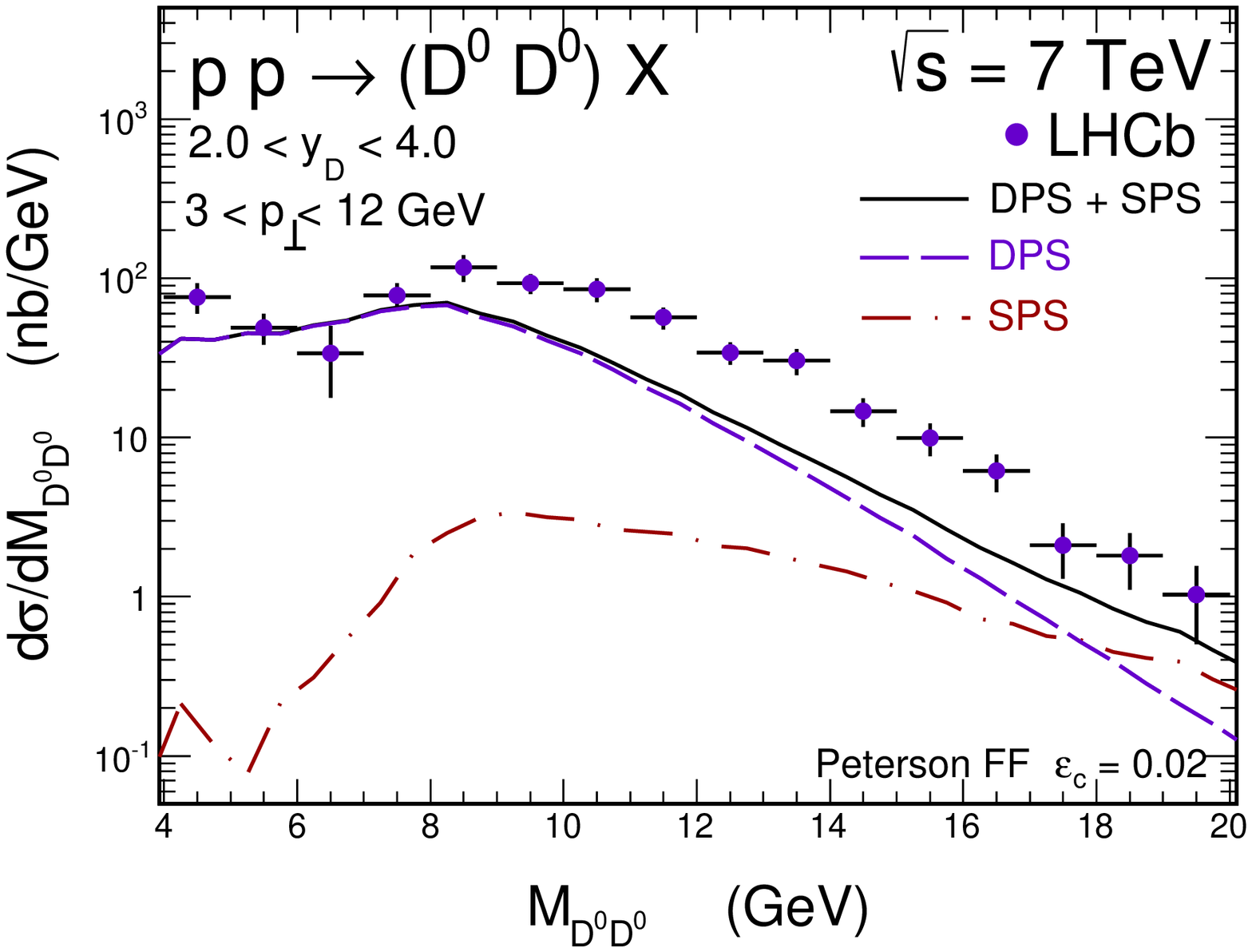}
   \caption{
\small Distributions in meson transverse momentum when both mesons are
measured within the LHCb acceptance and corresponding distribution
in meson-meson invariant mass for DPS and SPS contributions.
}
 \label{fig:Minv-and-pt-mesons}
\end{figure}

In Fig.~\ref{fig:Minv-and-pt-mesons} we show distributions in meson transverse momentum (left panel) and meson-meson invariant mass (right panel).
The shape in the transverse momentum is almost correct but some cross
section is lacking. In the case of the $DD$ invariant mass distribution one can also see some lacking strength at large invariant masses.
In both cases, the SPS contribution (dash-dotted line) is compared to the DPS one (dashed line). The dominance of the DPS mechanism in description of the LHCb double charm data is clearly confirmed. The DPS mechanism gives a sensible explanation of the measured
distribution, however some strength is still missing. This can be due to the single-ladder-splitting mechanisms discussed recently in Ref.~\cite{Gaunt2012}.

\begin{figure}[!h]
\centering
\includegraphics[width=6.cm]{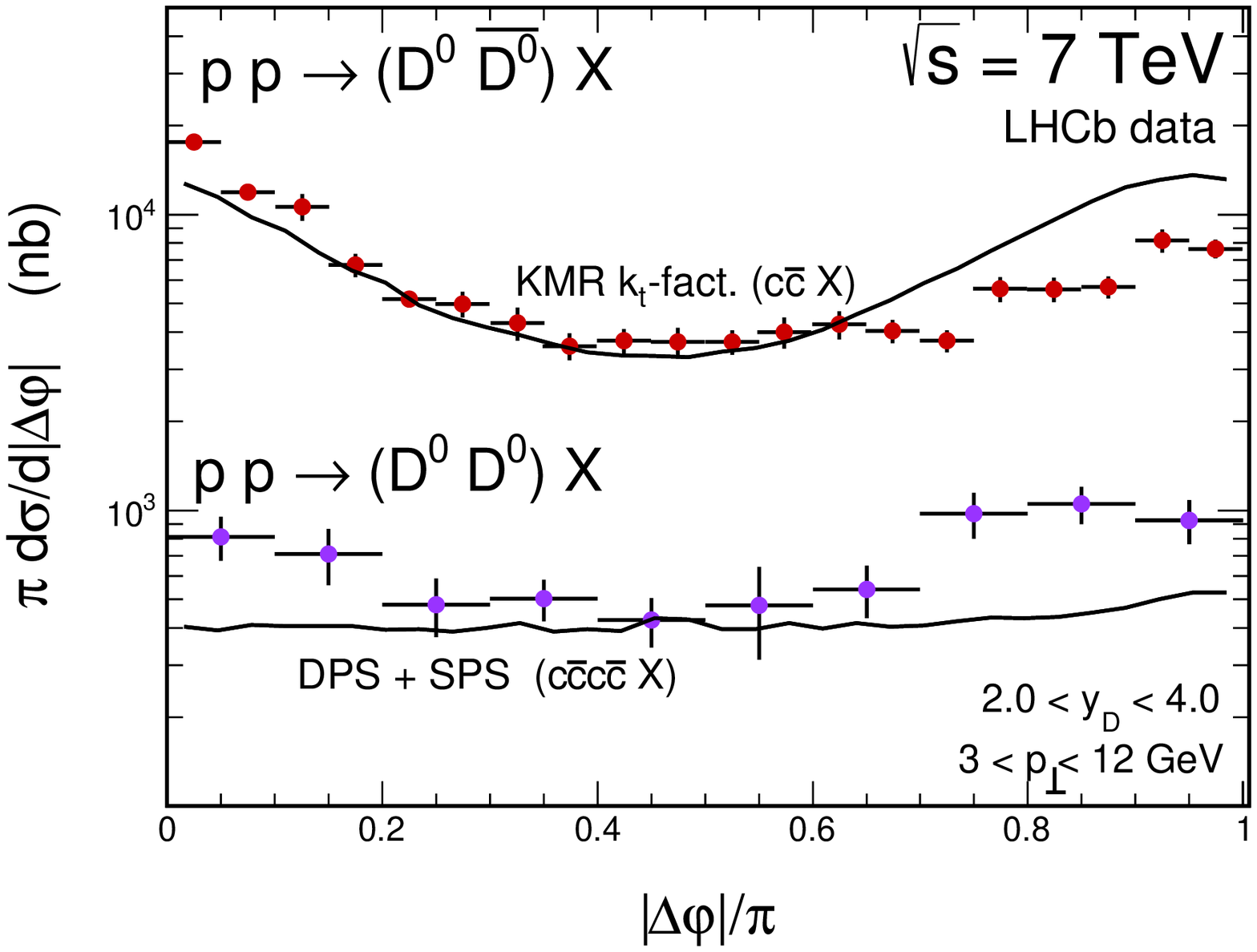}
\includegraphics[width=6.cm]{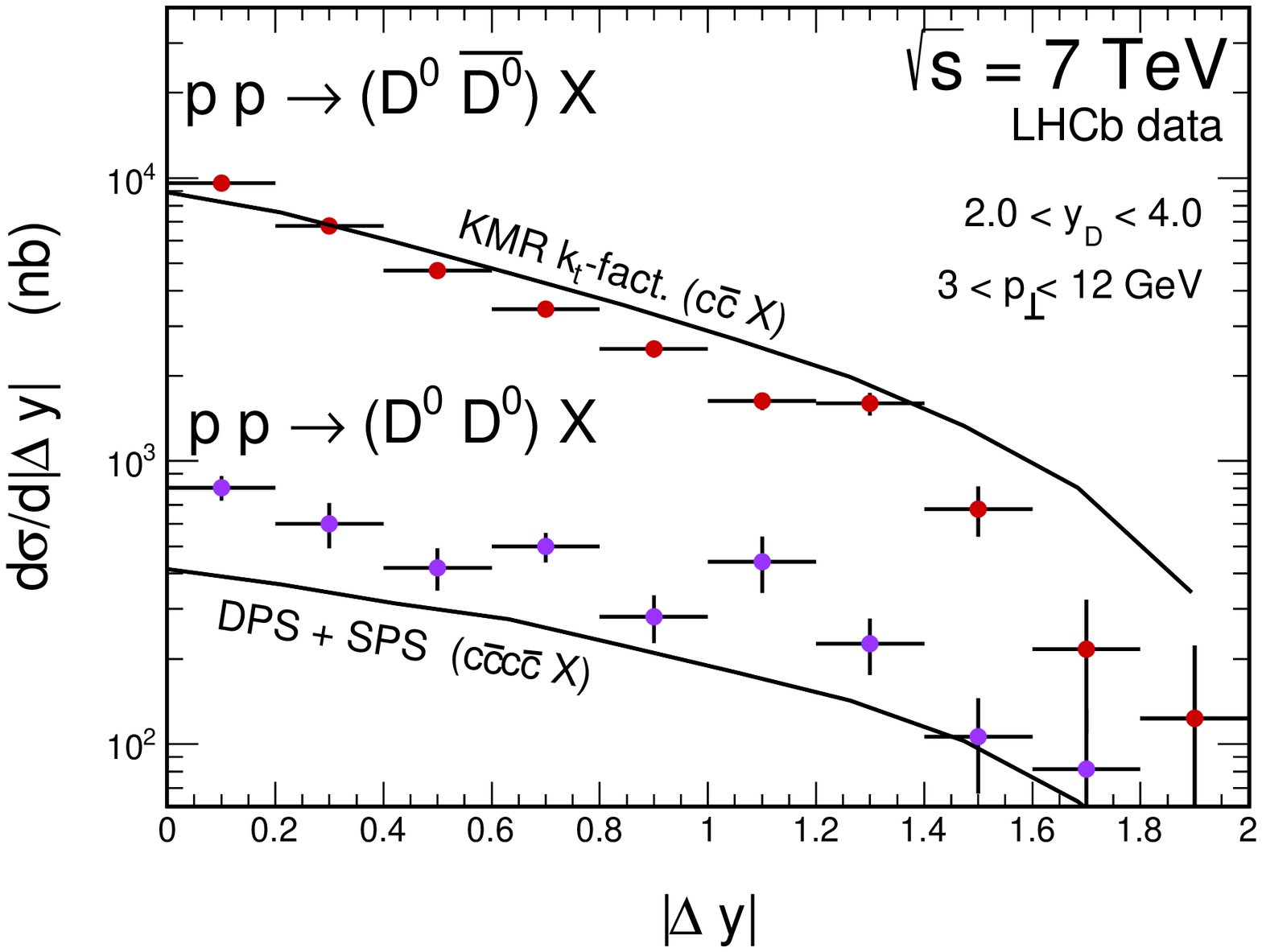}
   \caption{
\small Azimuthal angle correlation between $D^0 D^0$ and $D^0 \bar D^0$ (left) and distribution in rapidity distance between two $D^0$ mesons and
between $D^0 \bar D^0$.
}
 \label{fig:DPS-SPS}
\end{figure}

In Fig.~\ref{fig:DPS-SPS} we compare correlations for $D^0 D^0$ and 
$D^0 \bar D^0$ in azimuthal angle (left panel) and in rapidity distance (right panel).
In the case of the azimuthal angle distribution the experimental data suggest some small correlations at small and large angles
in contrast to the flat result of the standard DPS calculation.
In addition, the azimuthal angle distribution for identical mesons is somewhat
flatter than that for $D^0 \bar D^0$ which is consistent with the
claim of the dominance of the DPS contribution. The rapidity distance distribution for $D^0 \bar D^0$ falls down
somewhat faster than in the case of identical $D^0$ mesons.

Recently, it has been found that there are (at least) two different 
types of contribution to the DPS cross section, which are accompanied
by different geometrical prefactors ($\sigma_{eff,2v2}$ and $\sigma_{eff,2v1}$) (see e.g. \cite{GS2010}). One of these is the standard conventional contribution (2v2) in which two separate ladders emerge from both
protons and interact in two separate hard interactions (see the left panel of Fig.~\ref{fig:diagrams_ccbarccbar}).
This is the one that is often considered in phenomenological analyses and has been
applied in the studies presented above. 

The other type of process is the perturbative single-ladder-splitting (2v1)
which is similar to the conventional mechanism except that
one proton initially provides one ladder, which perturbatively
splits into two (see the middle and right diagrams in Fig.~\ref{fig:diagrams_ccbarccbar}).
Recently, the relative importance of the conventional and single-ladder-splitting DPS processes has been studied in Ref.~\cite{GMS2014}.

\begin{figure}[!h]
\centering
\includegraphics[width=6.cm]{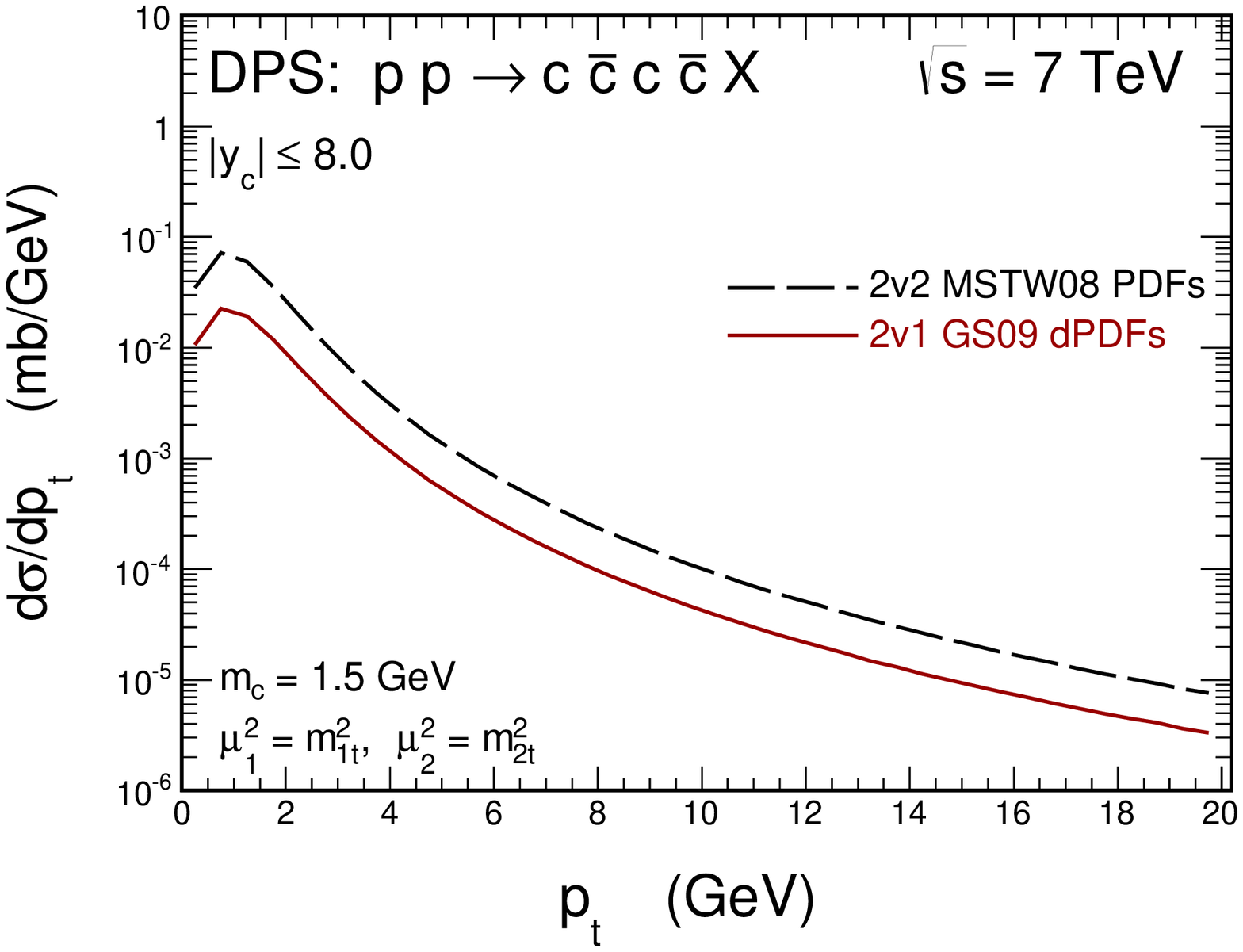}
\includegraphics[width=6.cm]{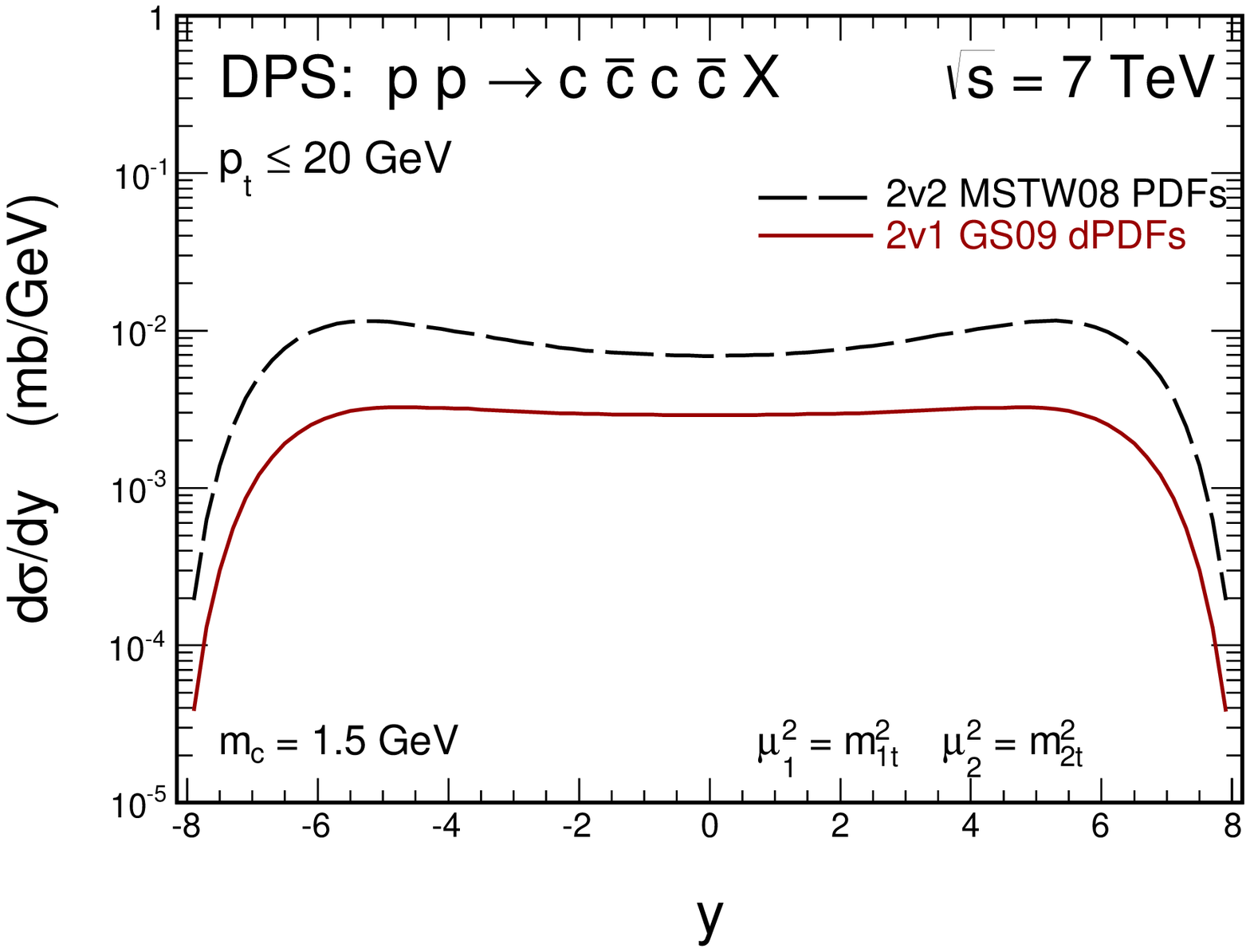}
   \caption{
\small Transverse momentum (left panel) and rapidity (right panel) distribution of charm quark/antiquark 
for $\sqrt{s}$ = 7 TeV for conventional and single-ladder-splitting DPS mechanisms. 
}
 \label{fig:dsig_dpt}
\end{figure}

In Fig.~\ref{fig:dsig_dpt} we show transverse momentum (left panel) and rapidity (right panel) distribution of the charm
quark/antiquark for the DPS mechanisms calculated within LO collinear approach at $\sqrt{s}$ = 7 TeV. 
The conventional and splitting terms are shown separately. 
The splitting contribution (lowest curve, red online) is smaller, 
but has almost the same shape as the conventional DPS contribution. 
The ratio of the DPS single-ladder splitting contribution to the conventional one in the case of double-charm production has been roughly estimated to be 
at the level of $30-60\%$. According to these results, the missing strengths in the description of the LHCb double charm data seem to be at least partially related to the 2v1 contribution.

The differential distributions in rapidity and in transverse momentum
for the conventional and the parton-splitting contributions have essentially the same shape. This makes their model-independent separation
extremely difficult. This also shows why the analyses performed are able to give reasonably good description
of different experimental data sets only in terms of the conventional DPS contribution. The
sum of the conventional and standard DPS contributions behaves almost exactly like the conventional contribution alone, albeit
with a smaller $\sigma_{eff}$ that depends only rather weakly on energy, scale and momentum
fractions.

\begin{figure}[!h]
\centering
\includegraphics[width=6cm]{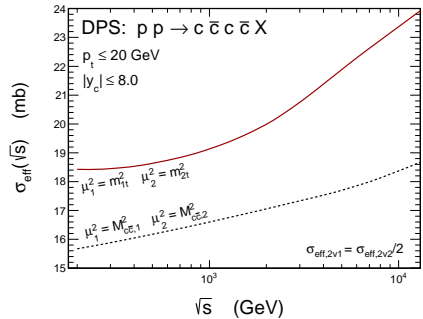}
   \caption{
\small Energy and factorization scale dependence of $\sigma_{eff}$
for $c \bar c c \bar c$ production as a consequence of existence
of the 2v2 and 2v1 components. 
In this calculation we have taken
$\sigma_{eff,2v2}$ = 30 mb and $\sigma_{eff,2v1}$ = 15 mb.
}
 \label{fig:sig_eff_charm}
\end{figure}

In Fig.~\ref{fig:sig_eff_charm} we show how the empirical (experimentally accessible)  
$\sigma_{eff}$ value depends on centre-of-mass energy when both 2v2 and 2v1 components are taken into account (assuming that 
the value of $\sigma_{eff,2v2}$ is independent of energy). 
We see a clear dependence of $\sigma_{eff}$ on energy in the plot, and also
on the renormalization/factorization scales. Assuming that there is no other mechanism
for an energy dependence, $\sigma_{eff}$ is 
therefore expected to increase with centre-of-mass energy. 
Note also that the empirical $\sigma_{eff}$ value obtained here is 
in the ballpark of the values extracted in experimental
measurements of DPS ($15-20$ mb).

It is not clear in the moment how to combine the higher-order effects with the perturbative splitting mechanism.
An interesting question is whether the ratio between the conventional and splitting contributions changes
when higher-order corrections are included. Further studies in this context are clearly needed to fully include the splitting DPS
contributions for the LHCb double charm experimental data.

%
%
%

\end{document}